\documentclass[aps,prd,showpacs,preprintnumbers,onecolumn]{revtex4}
\usepackage{epsfig}
\usepackage{latexsym,amsmath,amssymb,amstext}
\begin{document}
\title{A faster method of computation of lattice quark number susceptibilities}
\author{R.\ V.\ \surname{Gavai}}
\email{gavai@tifr.res.in}
\affiliation{Department of Theoretical Physics, Tata Institute of Fundamental
       Research, Homi Bhabha Road, Mumbai 400005, India.}
\author{Sayantan\ \surname{Sharma}}
\email{sayantan@physik.uni-bielefeld.de}
\affiliation{Fakult\"at f\"ur Physik, Universit\"at Bielefeld, 
           D-33615 Bielefeld, Germany}

\begin{abstract}
We compute the quark number susceptibilities in two flavor QCD for staggered
fermions by adding the chemical potential as a Lagrange multiplier for the
point-split number density term. Since lesser number of quark propagators are
required at any order, this method leads to faster computations.  We propose a
subtraction procedure to remove the inherent undesired lattice terms and check
that it works well by comparing our results with the existing ones where the
elimination of these terms is analytically guaranteed.  We also show that 
the ratios of susceptibilities are robust, opening a door for better estimates
of location of the QCD critical point through the computation of the tenth and
twelfth order baryon number susceptibilities without significant additional
computational overload. 
\end{abstract}
\preprint{BI-TP 2011/42, TIFR/TH/11-47} 
\maketitle
\section{Introduction}
Quantum Chromo Dynamics (QCD), the theory of strong interactions, may have a
critical point in the temperature (T)-baryon number density (or the baryonic
chemical potential $\mu_B$) depending on the number of light flavors it has.
The search for this critical point is one of the major experimental goals of
many heavy ion collider experiments world wide. In particular, the STAR
experiment at the Brookhaven National Laboratory has begun an extensive Beam
Energy Scan(BES) program~\cite{mohanty}.  It aims to scan the QCD phase diagram
for baryon chemical potential between 20 to around 400 MeV, looking for the
signatures of the presence of the critical point. FAIR at GSI, Darmstadt,
Germany and NICA at Dubna, Russia are proposed to be operational in future with
a key objective as the search for and the study of, a QCD critical point.  It,
therefore, appears natural to have a first principles theoretical exploration
of the QCD critical point, with a possible prediction for its location which
could serve as a useful reference for these large scale experiments. The
physics near the critical point is essentially non-perturbative. Lattice gauge
theory is the most successful non-perturbative tool which can be used for such
an exercise.  Fluctuations of conserved charges like the baryon number or
strangeness are sensitive indicators of the existence of singularities in the
phase diagram like the critical point~\cite{koch,srs}. It is expected that the
baryon number susceptibility would diverge at the critical point. On a finite
lattice, however, this quantity will show a peak at the critical point which
would get sharper as the lattice volume is increased.  The direct computation
of the susceptibility at finite baryon density is difficult due to the infamous
``fermion sign problem''. One of the techniques to circumvent the sign problem
at finite density is to compute the baryon number susceptibility as a Taylor
series expansion in chemical potential near zero~\cite {alt1,alt2,gg3}. The
radius of convergence of the series should yield an estimate of the location of
the critical point~\cite{gg1}. For the precise estimation of the radius of
convergence, one needs to compute ratios of as many higher orders of baryon
number susceptibilities as possible.  Current state of the art is the eighth
order susceptibility on the lattice~\cite{gg2}.  Higher order terms are
important in determining the critical point, and extending to higher orders in
Taylor series is therefore desirable although the explosion in the CPU time
required is severely constraining.

There are two major issues that have to be addressed for the efficient
computations of higher order quark number susceptibilities(QNS).  Firstly,
using the standard method of introducing chemical potential on the
lattice~\cite{hk,kogut,bilicgavai,gavai}, the computation of QNS beyond eighth
order with reasonably good precision becomes rather expensive. At each order,
the fermion matrix inversions account for the maximum time involved in
computing the QNS. As shown in~\cite{gg1}, twenty inversions are necessary for
computing the eighth order susceptibility; it would increase to about forty for the
tenth order. For computing higher orders of QNS on the lattice, the number of
matrix inversions would thus increase drastically thereby increasing the
computation time.  Secondly the current estimates of the susceptibilities
beyond fourth order on the lattice tend to become statistically demanding as
they are rather noisy. This is due to the fact that there are delicate
cancellations in the expressions of QNS between different terms.  Moreover, the
number of such terms itself increases progressively with the order of the
susceptibility. Each term has to be computed with appropriate precision in
order to ensure the cancellation of is free of computational artifacts. It
would be desirable to reduce the number of such cancellations.

In this paper we attempt to address both these issues. We use a staggered
fermion matrix in which the chemical potential, $\mu$, enters as a Lagrange
multiplier, multiplying the point-split conserved number density term.  It
alleviates the problems mentioned above, as shown in ~\cite{gs}.  For computing
the eighth order QNS, only eight fermion matrix inversions would be
needed~\cite{gs} as compared to the twenty required in the conventional
method~\cite{hk}.  In the early years of lattice computations this method of
introducing $\mu$ linearly was discarded because the second order QNS computed
using this term leads to a $1/a^2$ lattice term that diverges in the limit of
vanishing lattice spacing, $a$.  Modifications of the lattice
action~\cite{hk,kogut,bilicgavai,gavai} eliminated this term exactly for the
free theory.  In addition, they also ensured the correct Fermi surface on the
lattice.  These modifications lead to the difficulties in computing the
higher order QNS though. It therefore appears a worthy attempt to focus
back on the linear in $\mu$ action and devise ways to obtain physical answers.
A method of successful elimination of the divergence maybe relevant in another 
context as well.  The recently proposed overlap operator at finite 
density~\cite{ns} with exact chiral symmetry on the lattice has such a linear
dependence in $\mu$.   Indeed, the suggestions of 
~\cite{hk,kogut,bilicgavai,gavai} of divergence removal do not seem to work in
that case. 
 
Our paper is organized as follows. In Section~\ref{basic formalism} we briefly
review the basic formulae of the various susceptibilities and show the
differences in the two methods.  In Section~\ref{results} we suggest a
procedure to remove the lattice artifacts in the QNS computed with the linear
term in $\mu$. Free fermion results are shown to yield the correct continuum
limit with it.  We then compare our results for full QCD 
with those in the standard exponential method~\cite{hk}.  We
show that our results are consistent with the existing results for 
different orders of QNS and the ratios of baryon number susceptibilities for a 
wide range of temperatures.  A summary of our results along with possible sources of 
errors and their refinement is given in the last section.

\section{Formalism}
\label{basic formalism}
The quark number susceptibilities(QNS) for two flavor QCD are defined as,
\begin{equation}
\label{eqn:defsusc}
\chi_{ij}(\mu_u,\mu_d)=\frac{T}{V}\frac{\partial^{i+j} \ln Z(T,\mu_u, \mu_d,
,m_u, m_d)}
{\partial \mu_i\mu_j}
\end{equation}
where $Z$ is the QCD partition function,
\begin{equation}
\label{eqn:parfunc}
 Z(T,\mu_u,\mu_d)=\int \mathcal{D}U
 \rm{e}^{-S_G}Det D_u^{\frac{1}{4}}Det D_d^{\frac{1}{4}}.
\end{equation}
$S_G$ is the action for the gluon fields. We use the standard plaquette action 
for $S_G$.  $D_i$, for each $i=u,d$,  is the staggered fermion matrix in 
presence of finite quark chemical potentials.  Several choices of the Dirac 
matrix are possible on the lattice at finite density.  As in the continuum,
the chemical potential can be introduced as a Lagrange
multiplier corresponding to the conserved number density on the lattice in
the point split form.  This term is added to the standard staggered fermion
matrix to yield the $D(\mu)$ which we use in this work:
 \begin{eqnarray}
\nonumber
 D(\mu)_{xy}&=&D(0)_{xy} +\mu a\left[
\eta_4U^{\dagger}_4(y)\delta_{x,y+\hat{4}}+\eta_4 U_4(x)
\delta_{x,y-\hat{4}}\right] \\
            &=& \sum_{i=1}^3 \left[
\eta_i U_i(x)\delta_{x,y-\hat{i}}-\eta_iU^{\dagger}_i(y)\delta_{x,y+\hat{i}}\right] - 
(1 -\mu a)\eta_4U^{\dagger}_4(y)\delta_{x,y+\hat{4}}+ (1 +\mu a)
\eta_4 U_4(x) \delta_{x,y-\hat{4}} + m a~ \delta_{x,y}.
\label{eqn:diracop}
\end{eqnarray}
$\eta_i$ are the phase factors remnants of the gamma matrices as usual. 
Replacing $( 1\pm \mu a)$ by exp$(\pm \mu a)$, one obtains the popular
method~\cite{hk} of introducing $\mu$.  We shall compare our results
with those obtained by employing the latter.

In general, the chemical potentials $\mu_i$ and $\mu_j$, for quark flavors i
and j respectively need not be the same. But since isospin is a good symmetry
for QCD, we set the chemical potentials for up and down quarks to be the same,
$\mu_u=\mu_d=\mu$. Hence the baryon chemical potential is just $\mu_B=3\mu$.  
The baryon number susceptibilities can then be expressed in terms of the quark
number susceptibilities(QNS) $\chi_{ij}$. For two flavor QCD, the expressions
for baryon number susceptibility of n-th order, $\chi_{B}^n$, are 
\begin{eqnarray}
\label{eqn:chib}
\nonumber
 \chi_{B}^{(4)}&=&\frac{1}{2}\left[\chi_{40}+2\chi_{31}+\chi_{22}\right]~,~\\\nonumber
\chi_{B}^{(6)}&=&\frac{1}{4!}\left[\chi_{60}+4\chi_{51}+7\chi_{42}+4\chi_{33}\right]~,~\\
\chi_{B}^{(8)}&=&\frac{1}{6!}\left[\chi_{80}+6\chi_{71}+16\chi_{62}+26\chi_{53}+15
\chi_{44}\right]~.
\end{eqnarray}
In this work we would be interested in computing the baryon number
susceptibilities at $\mu_B=0$ since these quantities appear in the Taylor
series expansion of the second order baryon number susceptibility expressed in
powers of $\mu_B$,
\begin{equation}
 \label{eqn:chi2ts}
\frac{\chi_{20}(\mu_B)}{T^2}=\frac{\chi_{20}(0)}{T^2}+\chi^{(4)}_{B}
\left(\frac{\mu_B}{3T}\right)^2+\chi^{(6)}_{B}\left(\frac{\mu_B}{3T}\right)^4
+\chi^{(8)}_{B}\left(\frac{\mu_B}{3T}\right)^6+..
\end{equation}
If a critical point exist then the baryon number susceptibility should 
diverge at that point in the continuum limit. The radius of convergence of this
series, should therefore determine the location of the critical point in the 
$T$-$\mu_B$ plane of the QCD phase diagram.  It can be defined as
\begin{equation}
\label{eqn:radc}
r_n=\lim_{n\rightarrow\infty} \sqrt{\frac{\chi_B^{(n+1)}}{\chi_B^{(n+3)}}}~,~
{~~~\rm or~~~}
r_n=\lim_{n\rightarrow\infty} \left[\frac{\chi_B^{(2)}}{\chi_B^{(n+2)}}\right]^{1/n}~.
\end{equation}
At the critical point, all the $\chi_B^{(n)}$ are positive definite and the
successive estimates of the radius of convergence agree with each
other~\cite{gg1}. It is clear from  the above definitions of the radius of
convergence that one needs to estimate more and more higher orders of the 
baryon number susceptibilities in order to locate it with precision and
reliability.

The QNS in Eq. (\ref{eqn:defsusc}) can be written in terms of the trace of the
derivatives of the Dirac matrix in Eq. (\ref{eqn:diracop}) and its inverse at
$\mu=0$. All the expressions for the QNS upto eighth order are given in the
Appendix of Ref.\cite{gg1} as expectation values of $\mathcal{O}_{ijkl..}$.
These in turn are written in terms of the derivatives and inverse of $D$.  We
note that as a consequence of changing to our lattice Dirac operator linear in
chemical potential, only the expressions for $\mathcal{O}_{ijkl..}$ change,
and indeed simplify a lot. This is because the  second and higher order
derivatives with respect to the chemical potential of the $D$ in Eq.
(\ref{eqn:diracop}) vanish.  For example, the expression of $\chi_{40}$ in our
case in the notation of Ref. \cite{gg1} is still,
\begin{equation}
 \chi_{40}=\frac{T}{V}\left[\langle\mathcal{O}_{1111}+6\mathcal{O}_{112}+4\mathcal{O}_{13}
+3\mathcal{O}_{22}+\mathcal{O}_{4}\rangle-3\langle \mathcal{O}_{11}+\mathcal{O}_{2}\rangle^2\right]~,~
\end{equation}
with each such $\mathcal{O}_{n}$ simply given by 
\begin{equation}
\mathcal{O}_{n}=(-1)^{n-1} (n-1)!~\text{Tr}(D^{-1}D')^n~. 
\end{equation}
In order to appreciate the difference, let us point out as an example
$\mathcal{O}_{2}$ and $\mathcal{O}_{4}$ in our case are:
\begin{equation}
\mathcal{O}_{2}=-~\text{Tr}(D^{-1}D')^2~ {~~\rm and ~~} 
\mathcal{O}_{4}=-~ 6 \text{Tr}(D^{-1}D')^4~,~ 
\label{eqn:uso2o4}
\end{equation}
while in Ref. \cite{gg1} they are
\begin{eqnarray}
\label{eqn:gko2}
\nonumber
\mathcal{O}_{2}&=&-~\text{Tr}(D^{-1}D')^2 + \text{Tr}(D^{-1}D'') 
~ {~~\rm and ~~} \\
\mathcal{O}_{4}&=&-6\text{Tr}(D^{-1}D')^4 + 12\text{Tr}[(D^{-1}D')^2 D^{-1}D'']
-3\text{Tr}(D^{-1}D'')^2 -4\text{Tr}(D^{-1}D'D^{-1}D''')+ \text{Tr}(D^{-1}D'''') ~.
\label{eqn:gko4}
\end{eqnarray}

From a comparison of the Eq. (\ref{eqn:uso2o4}) and Eq. (\ref{eqn:gko2}), one
sees the absence of the second derivative term in the former.  In fact, it
arises due to those modifications which eliminate the free theory divergence.
But then due to the same reasons Eq. (\ref{eqn:gko4}) has four additional terms
of alternating sign as compared to Eq. (\ref{eqn:uso2o4}).  This is generic.
As the order n increases, the number of such terms in the expression of the
$\mathcal{O}_{n}$ increases.  It should be noted that each trace in the
expressions of QNS contains product of the inverse of the Dirac operator with
the derivatives of Dirac operator with respect to $\mu$. The matrix inversion
is the most expensive computation on the lattice. If successive derivatives of
Dirac operator are all finite, one has to compute more number of matrix
inversions.  Using the operator defined in Eq.  (\ref{eqn:diracop}) only the
first derivative of the Dirac operator is finite.  Thus the price of additional
terms with sign changes increases at the higher orders.  Computationally, this
implies more inverses of the matrix $D$ and more precision with each term has
to be computed in order to get $\mathcal{O}_n$ with comparable accuracy for all
$n$.   It therefore seems a worthy effort to devise a scheme to remove the free
theory divergence in another way, as we attempt in this work.  In one can
successfully remove the lattice artifacts that appear in the expressions of the
lower order susceptibilities, then it can potentially open the door to compute
the eighth and the higher derivatives with considerably less computational
effort, enabling a better check on the radius of convergence estimate, as
argued in~\cite{gs}.

\section{Results}
\label{results}
In this section we first begin by computing the susceptibilities of free fermions 
using the staggered operator in Eq. (\ref{eqn:diracop}) in order to 
separate the artifacts in the second and fourth order QNS. 
In~\cite{gs} we proposed to remove the undesired artifacts by estimating the
zero temperature contribution on a symmetric lattice for each value of the
gauge coupling, as is done for the pressure or energy density computation.
This was motivated by the idea of keeping the cut-off effects to be the
same as temporal lattice size increases.   On the other hand, if one notes
that the all the analytic methods of removal of the $\mu^2$-divergent terms
were proven to be so only for the free theory, and that no new divergences
appeared in the simulations of the interacting theory, one is lead to 
believe that a numerical method devised also for the free theory may work
as well.  Furthermore, since no new renormalizations are necessary at finite 
$T$ and $\mu_B$ in perturbation theory, we expect the free theory artifacts to be the 
dominant ones.  In this work we show that this method of
subtraction gives results for all QNS which are in good agreement with the 
existing results for $T>T_c$.  For $T<T_c$, this subtraction method appears 
to lead to some differences but these are small enough to be tolerated as
differences in the finite parts of two schemes of removal of infinity.  On
finer lattices these will shrink, if this is indeed so.   Moreover, even
for $T<T_c$, we find that the crucial ratios of susceptibilities and
therefore, the radius of convergence estimates are not sensitive to it.

\subsection{Free theory}
\begin{figure}
\begin{center}
\includegraphics[scale=0.6]{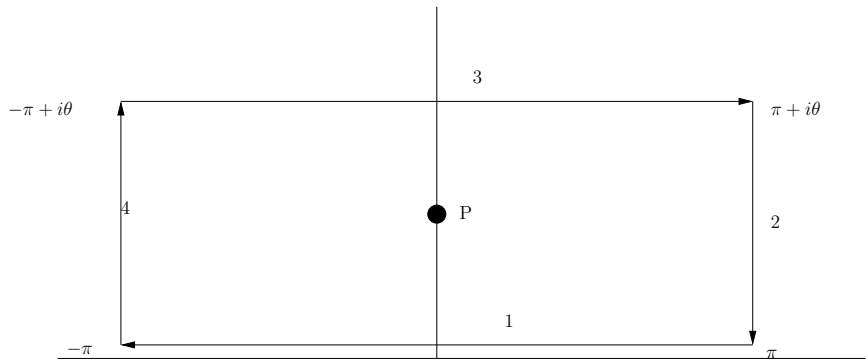}
\caption{The contour diagram for calculating the number density for free fermions at 
zero temperature on the lattice. P denotes the pole at 
$i\sinh^{-1}\frac{\sqrt{f}}{(1-\mu^2 a^2)^{1/4}}$.}
\label{contour}
\end{center}
\end{figure}

The number density and the quark number susceptibilities(QNS) can be calculated
analytically for the free fermions. We sketch how it is done, and discuss the
results for number density for free fermions, obtained with the fermion
matrix in Eq.  (\ref{eqn:diracop}), and the popular exponential form.  We
consider a lattice with $N$ sites along each spatial direction and $N_T$ sites
along the temporal direction. The lattice spacing is taken to be $a$ and
therefore the volume is $N^3 a^3$ and the temperature of the system is
$T=1/(N_T a)$. From Eq. (\ref{eqn:defsusc}), the expression for the number
density on the lattice is  
\begin{equation}
 \label{eqn:nodensity}
 na^3=\frac{i}{ N^3 N_T}\sum_{\vec p,n}\frac{(\sin\omega_n+i\mu a\cos\omega_n)\cos\omega_n}
 {f+(\sin\omega_n+i\mu a\cos\omega_n)^2}~ \equiv \frac{ i}{ N^3 N_T}\sum_{\vec p,\omega_n}
 F(\omega_n,\mu a,\vec p)~.
\end{equation}
where $f=(ma)^2+\sin^2(a p_1)+\sin^2(a p_2)+\sin^2(a p_3)$.  For the 
other form, $\mu$ appears only as $(\omega_n -i \mu a)$ in place of
$\omega_n$ above. 
The expression can be evaluated by the usual trick of converting the sum over 
energy states to a contour integral. 
The zero temperature limit for a fixed lattice spacing corresponds to
, $N_T\rightarrow\infty$. The energy eigenvalues then become 
continuous and lie in the range $[\frac{-\pi}{a},\frac{\pi}{a}]$. 
The expression for the number density in this limit is,
\begin{equation}
\label{eqn:ncont}
 na^3 = \frac{i}{ N^3}\sum_{\vec p}\left[-i F(\vec p,\mu a)\Theta
\left(\theta-\sinh^{-1}\frac{\sqrt{f}}{(1-\mu^2 a^2)^{1/4}}\right) +\int_{-\pi}^{\pi}
 \frac{d\omega}{2\pi}\frac{\sin\omega(\cos\omega+i\mu a\sin\omega)}
 {f+(1-\mu^2 a^2)\sin^2\omega}\right]~,~
\end{equation}
where $F(\vec p,\mu a)$ is the residue of the function $F(\omega-i\theta)$ in the 
positive half plane and $\tan\theta=a \mu$. In terms of the 
contour diagram in the complex $\omega$ plane in Fig. (\ref{contour}), the
original term is the line integral 3.  Completion of the contour leads to
the residue and since the contributions of the line integrals 2 and 4 cancel 
with each other, one is left with the contribution of the line integral 1.
It gives rise to additional lattice artifacts in the expressions of number 
density.  It vanishes for the exponential form due to $\omega \to - \omega$
symmetry.  Note, however, that the residue is present in both cases, and
leads to terms higher order in $a$ in each of them.
More specifically, in the continuum limit of $a \to 0$, the number density in 
the two cases are respectively 
\begin{eqnarray}
\nonumber
 n &\approx& \frac{2 \left(1-\sqrt{2}\right)\mu}{3 \pi ^2 a^2}+
\frac{\mu^3}{3\pi^2}\left[1-\frac{3\left(8-5
 \sqrt{2}\right)}{4}\right]+\frac{\mu^5a^2}{6\pi^2}\left[1-\frac{6\left(100-
\frac{125}{\sqrt{2} }\right)}{ 120}\right]+.. \\
    & =& -\frac{0.276\mu}{ \pi ^2 a^2}+
\frac{0.303\mu^3}{3\pi^2}+\frac{0.42\mu^5a^2}{6\pi^2}+..
\label{eqn:ngs}
\end{eqnarray}
and
\begin{equation}
\label{eqn:nhk}
 n=\frac{\mu^3}{3\pi^2}+\frac{\mu^5a^2}{6\pi^2}+..
\end{equation}  
Comparing the expressions in Eqs. (\ref{eqn:ngs}) and (\ref{eqn:nhk}), one
notices that the free theory divergence which, as mentioned earlier, exists
for the Dirac matrix in Eq.  (\ref{eqn:diracop}).  It is eliminated, on the
contrary, analytically in the latter case.    There is an additional
difference which matters in any physical comparison.  The second term of Eq.
(\ref{eqn:ncont}) contributes to the $\mu^3$ term in Eq. (\ref{eqn:ngs}) as
well, and reduces its value.  On the other hand, such a Fermi surface violation is
also cured in the usual method giving the correct coefficient for the $\mu^3$.
The expressions are similar in form for higher order terms which affect the
values of the sixth and higher order susceptibilities on a finite lattice.  
Indeed, the coefficient of the $\mu^5$ term is even larger in in Eq. 
(\ref{eqn:nhk}), a trend which persists for even higher terms as well.
Thus their approach in the continuum limit will be correspondingly slower. 

\begin{figure}
\begin{center}
\includegraphics[scale=0.6]{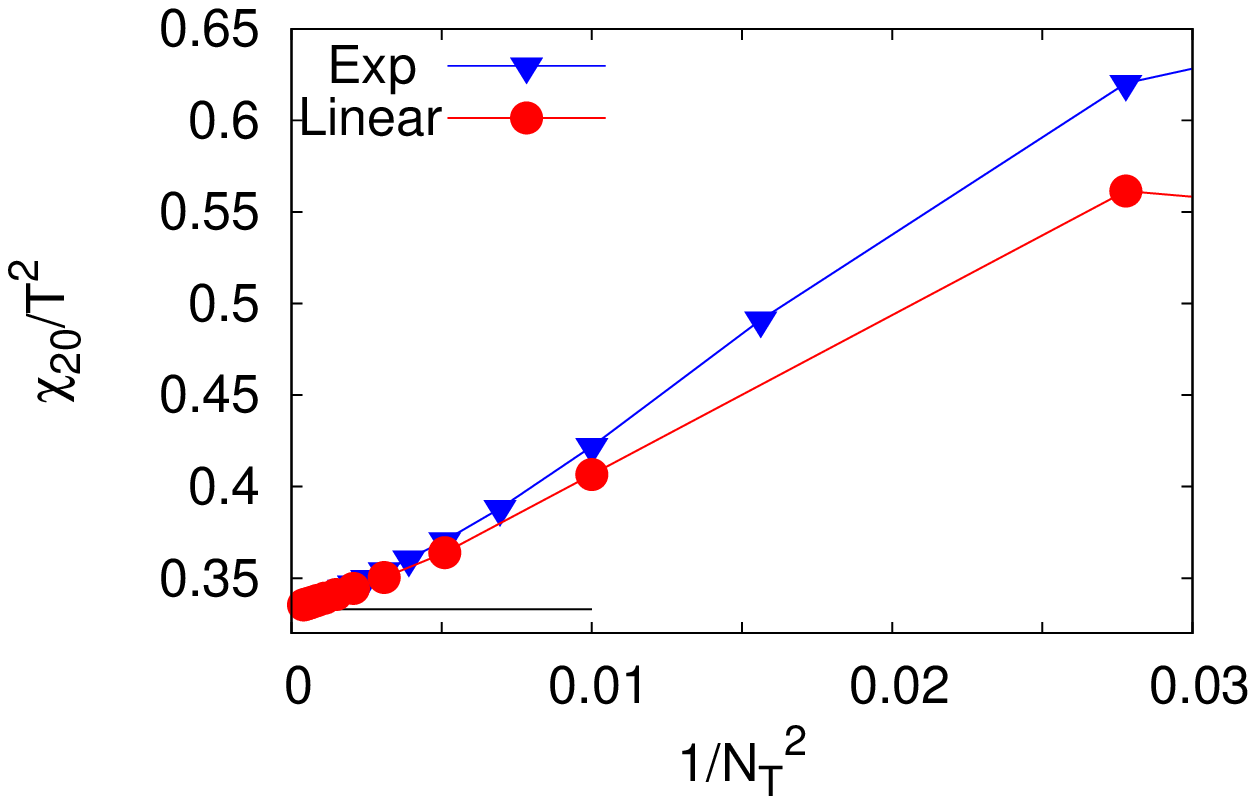}
\includegraphics[scale=0.6]{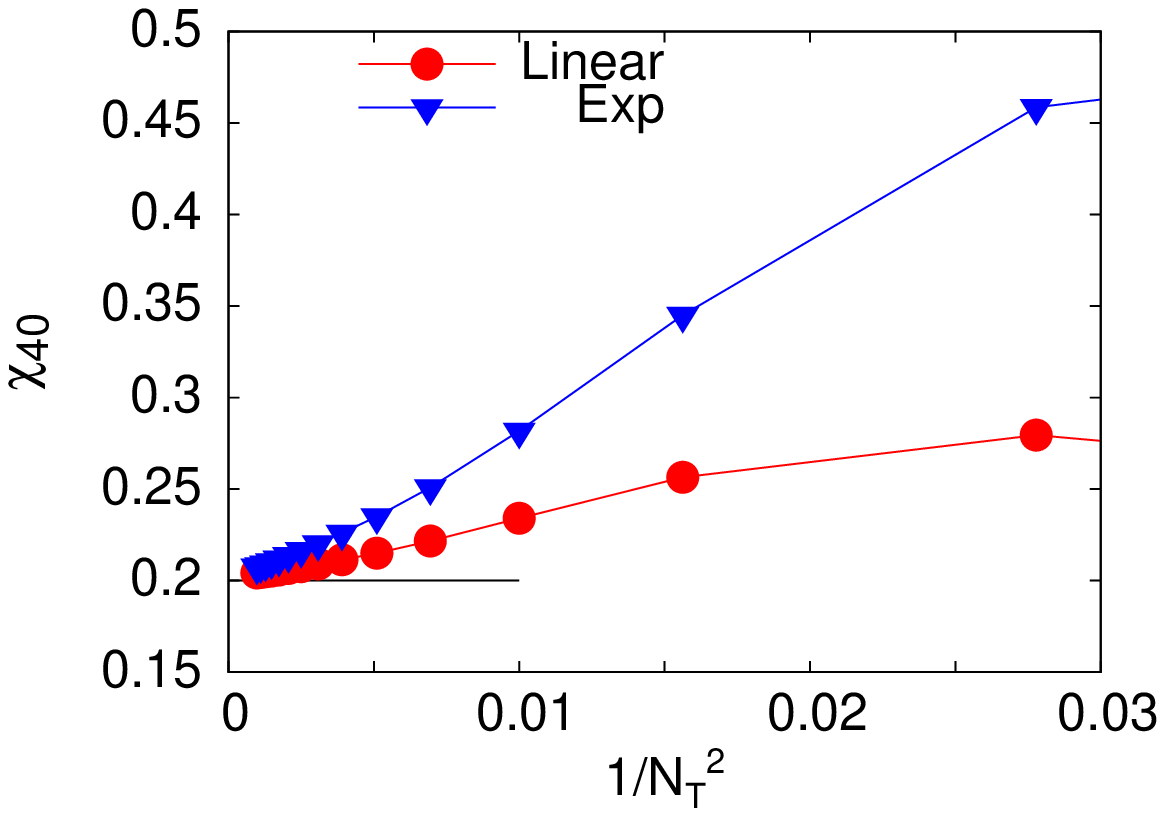}
\caption{The second (left panel) and the fourth order (right panel)
susceptibilities for free fermions as a function  $1/N_T^2$ for 
different methods of introducing $\mu$.}                                                                                
\label{freesusc24}
\end{center}
\end{figure}

\begin{figure}
\begin{center}
\includegraphics[scale=0.6]{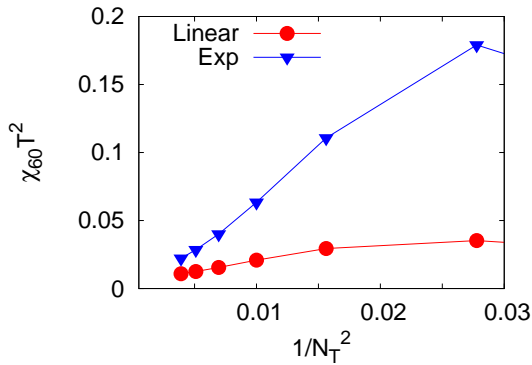}
\caption{The sixth order susceptibility for free fermions 
as a function  $1/N_T^2$ for different methods.}
\label{freesusc6}
\end{center}
\end{figure}

Since the QNS $\chi_{n0}(\mu=0)$ of interest to us here are computed from the
number density in Eq. (\ref{eqn:ngs}), they too have these lattice artifacts in
the form $\mathcal{O}(a^{n-4})$.  As we saw above, for $n \ge 6$ they exist for
both forms of the Dirac matrix, and are a bit less of a nuisance for the linear
form than the exponential one.  But, the $\chi_{20}$ has an
$\mathcal{O}(1/a^2)$ term which diverges in the continuum limit. It has to be
subtracted before the continuum limit is taken.  There is an additional
$\mathcal{O}(a^0)$ term in the expression for the fourth order susceptibility
which gives an incorrect result.  Clearly, removal of these artifacts is 
necessary.  Since we can trace them to the second term in 
Eq. (\ref{eqn:ncont}),  we propose to compute them directly numerically
from it, and subtract.  Thus taking one more derivative with respect to
$\mu a$, setting $a \mu = 0$ in the resultant expression, the subtraction
term of $\chi_{20}$ can be computed on a lattice of volume $N^3$ and infinite 
temporal extent by first analytically integrating over the energy eigenvalues 
along the temporal direction.   Thus we numerically evaluate 
\begin{equation}
 \label{eqn:chi2f}
\chi_{20}(0)=-\frac{1}{4~N^3}\sum_{\vec p}\left(1-\sqrt{\frac{f}{1+f}}\right)~,
\end{equation}
and
\begin{equation}
 \label{eqn:chi4f}
\chi_{40}(0)=-\frac{3}{4~N^3}\sum_{\vec p}\left(2-\frac{3+2f}{1+f}
\sqrt{\frac{f}{1+f}}\right)~.
\end{equation}
as the vacuum subtractions at zero temperature.  Eliminating these artifacts, 
the second and the fourth order susceptibilities 
are shown as a function of $1/N_T^2$ in Fig. (\ref{freesusc24}) and compared 
with the corresponding results obtained using the exponential form. In 
these plots the aspect ratio $N/N_T$ is fixed to be four as it yields already 
the thermodynamic limit. As $N_T$ becomes larger, corresponding to the continuum
limit, it is evident from the plots that our proposal for the 
artifact subtraction does ensure that these quantities indeed do have the 
correct continuum limit. 
The difference between the results for lattice sizes $N_T\leq 10$ are due to
finite cut-off effects which are comparatively larger for the H-K method. This
difference becomes more for the higher order QNS as shown in the
plot of the sixth order susceptibility in Fig. (\ref{freesusc6}).   

\subsection{Interacting theory}

In this section we compute the nth order baryon number susceptibilities at
vanishing $\mu_B$ for two flavor QCD using staggered fermions using the
operator given in Eq. (\ref{eqn:diracop}) and the subtraction scheme explained
above for the free theory.  Note that we follow all the existing computations in
this respect: the analytic cancellation of the divergence was shown only for
the free theory and the same form used for the interacting theory.   We used
the same configurations used previously for estimating all susceptibilities up
to the eighth order on a $N_T=6$ lattice~\cite{gg2}. For details of the
configurations and the scale setting, we refer the reader to the
Ref.~\cite{gg2}.  The lattice size used was $24^3\times6$ and the pion mass was
fixed at $M_{\pi}=230$MeV, as there. The critical coupling was defined from 
the peak of the unrenormalized Polyakov loop susceptibility.
 As discussed in the Section~\ref{basic formalism}, the 
expressions of susceptibilities contain trace of fermion
operator insertions.  For computing each trace, 500 random vectors were used.
This was also the optimum number of random vectors used in the earlier
work~\cite{gg1} for computation of the sixth and eighth order susceptibilities.
Thus, essentially all the computational details were maintained to be the
same as in~\cite{gg2}.  Our expressions for the various $\mathcal{O}_n$
are different and we use the subtraction scheme for $\chi_{20}$ and $\chi_{40}$.
Our results for QNS are compared with the corresponding results of
~\cite{gg2}, i.e., the exponential form and with analytic cancellation of the
divergence for the free theory.  Further, we also compute 
the ratios of our susceptibilities ans compare with the known results.

\subsection{Second order} 
\begin{figure}
\begin{center}
\includegraphics[scale=0.6]{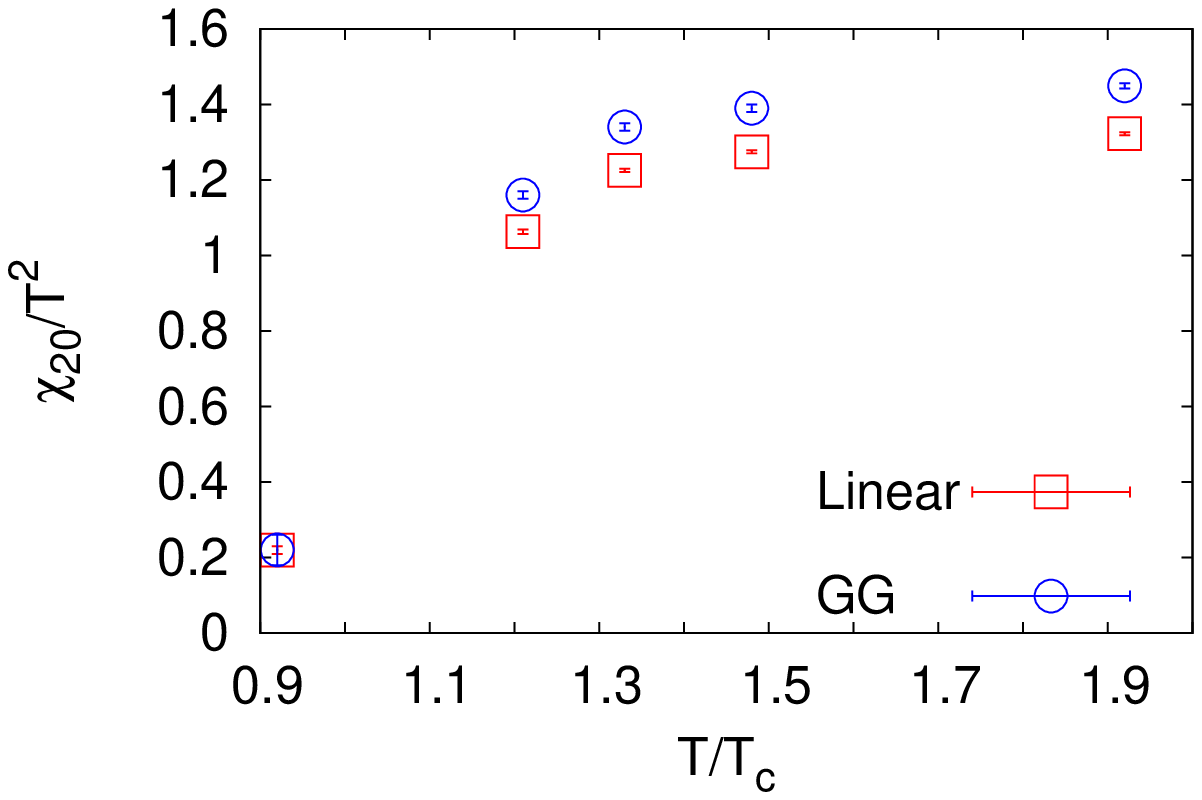}
\includegraphics[scale=0.6]{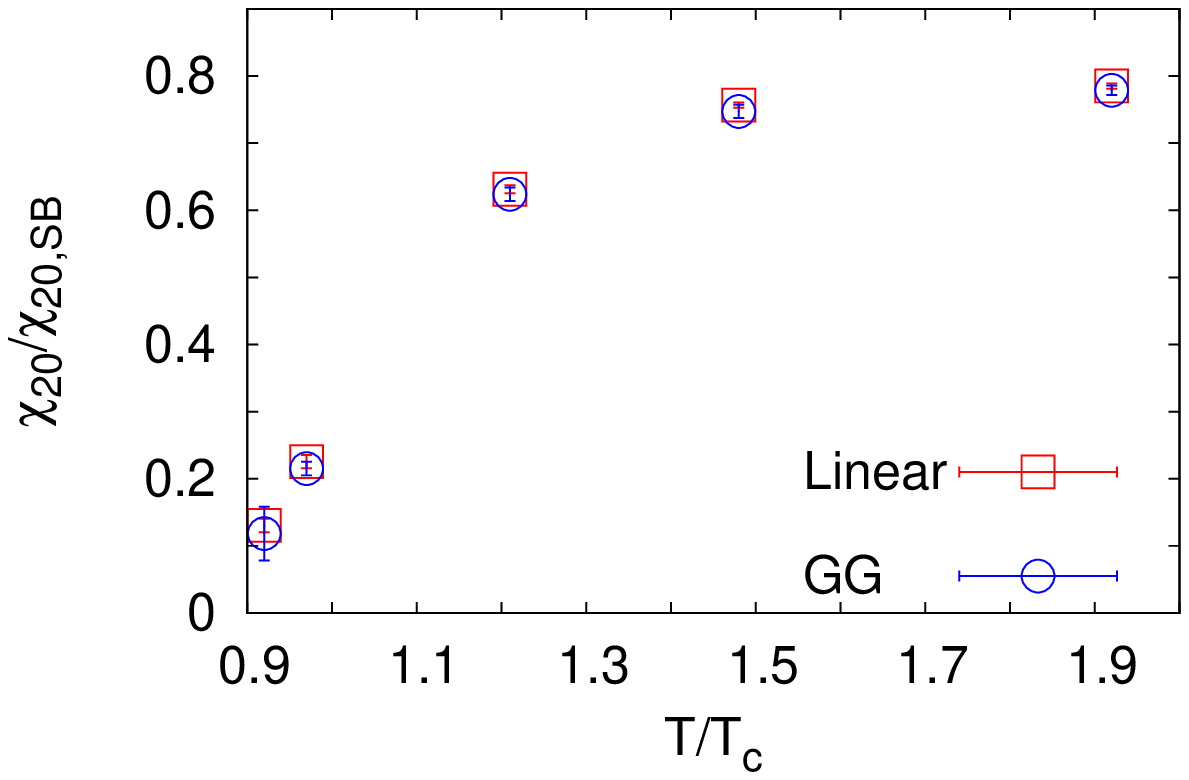}
\caption{The second order flavor diagonal(left panel)$\chi_{20}/T^2$ and the 
corresponding one normalized by the Stefan-Boltzmann value on a $N_T=6$ lattice 
(right panel) as a function of $T/T_c$.}
\label{chi20}
\end{center}
\end{figure}

We compute the zero temperature value of $\chi_{20}$ for free fermions using Eq. 
(\ref{eqn:chi2f}), by numerically summing over the momentum modes on a $24^3$
lattice since the spatial volume for our interacting case is $24^3$. We then
subtract this quantity from the $\chi_{20}$ values in the interacting theory 
computed using the $\mathcal{O}_2$ for our case.  The left panel of Fig. 
(\ref{chi20}) compares our results with those of ~\cite{gg2}, labeled
as `GG'.
At $0.92T_c$, the value of our baryon number susceptibility 
 matches with the existing GG results within errors. This suggests that including
our subtracted value the of $\langle \mathcal O_2 \rangle$ is 
effectively the same. Since the difference in the two cases comes only
from the Tr$~\left(M^{-1}M''\right)$ for ~\cite{gg2} and the subtraction
term of Eq.  (\ref{eqn:chi2f}) on a $24^3$ lattice for our case,
we can infer that their values are equal within the numerical 
precisions. Since one is evaluated in the interacting theory while the other
in the free theory, this equality justifies our {\em ansatz} that interactions
do not lead to further divergent terms, as expected also in perturbation theory.
Once the divergent term is subtracted the difference between the results 
in the two methods are due to different cut-off effects. The disagreement 
at the highest temperatures is consistent with the fact that there are
different cut-off effects in the free theory limit i.e at asymptotically high 
temperatures. They would vanish only in the continuum limit or very large
lattices.  Again this can be easily verified by numerical computations in
the free case.

We therefore check the variation of the ratio $\chi_{20}/\chi_{20,SB}$ computed
as a function of temperature where the Stefan-Boltzmann(SB) value is computed
on a $N_T=6$ lattice for both the cases. We expect that if the cut-off effects
in the interacting theory are not very different from the free theory for
temperatures much larger than $T_c$, this ratio should be independent of
the subtraction procedure used. 
From the right panel of Fig.  (\ref{chi20}) it is evident that the GG results
of ~\cite{gg2} are consistent with ours for $T>T_c$. At
$1.92~T_c$, the deviation from the Stefan-Boltzmann value is about the
same($\sim 20\%$) for both the methods.

\subsection{Fourth order}
\begin{figure}
\begin{center}
\includegraphics[scale=0.6]{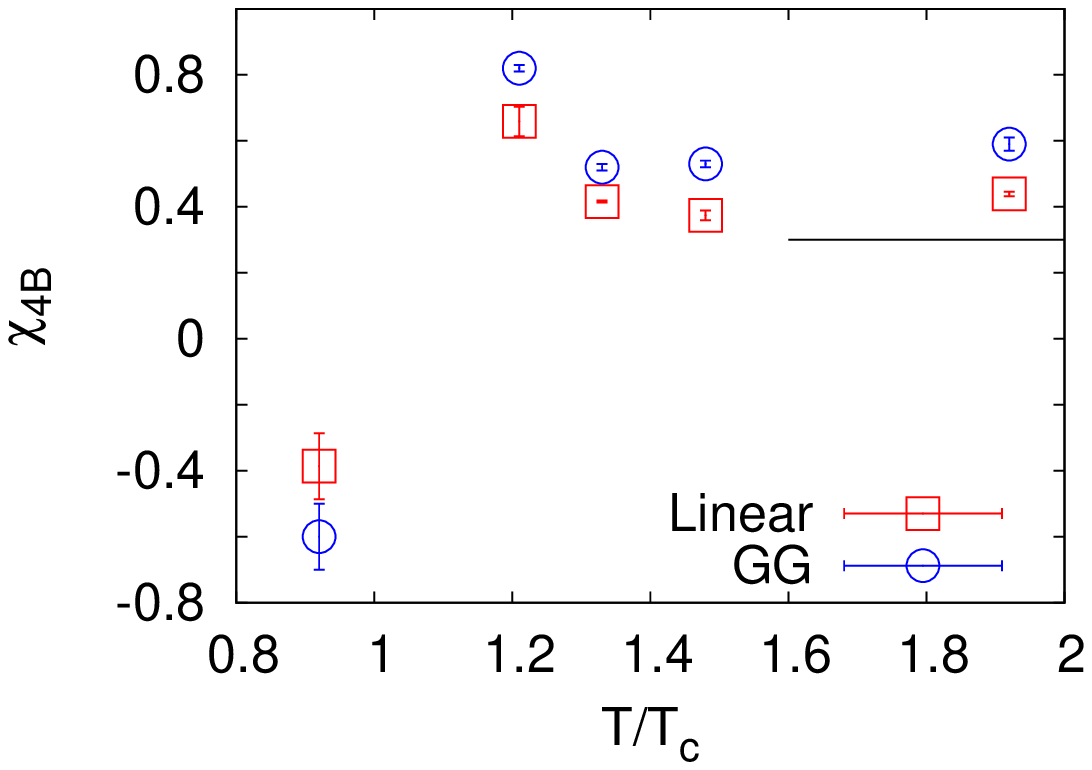}
\includegraphics[scale=0.6]{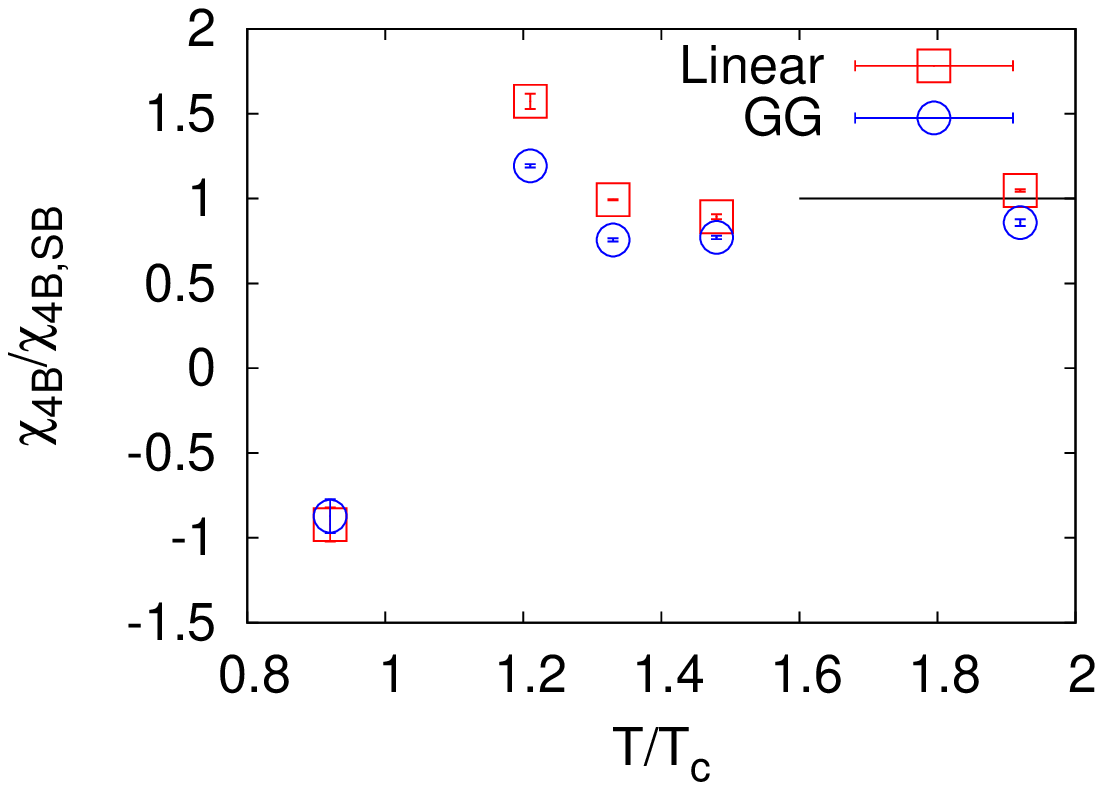}
\caption{The fourth order baryon number susceptibility(left panel) and the 
same quantity normalized by the Stefan-Boltzmann value on $N_T=6$ lattice  
(right panel) as a function of $T/T_c$.}
\label{chi4B}
\end{center}
\end{figure}

We follow the same subtraction procedure again for the fourth order
susceptibility and compute the subtraction constant using the expression in Eq.
(\ref{eqn:chi4f}) on a $24^3$ lattice.  Comparing the  $\chi_{4B}$ in the two
cases in the left panel of Fig.  (\ref{chi4B}), a good agreement is evident
with the existing GG results.  One does see, however, differences arising
perhaps due to the finite size of cut-off $a = 1/6T$.  In the right panel of
Fig. (\ref{chi4B}), the ratio of $\chi_{4B}$ and the corresponding Stefan
Boltzmann value on a $N_T=6$ lattice is shown as a function of $T/T_c$.  The
deviations from the continuum Stefan Boltzmann value are consistent with the
free theory results for $T>1.5~T_c$ indicating that the ideal gas cut-off
effects are likely to be more dominant than the interaction effects.  
At $1.92~T_c$, the ratio is away from unity by $\sim5\%$ in our computation
whereas the deviation is about $\sim15\%$ for the GG results.   
At $T=1.21~T_c$, the deviations are larger than that for free massless
fermions indicating larger effects due to interactions in the medium. 
Of course, continuum extrapolation are needed to make prediction on the nature
of the QCD medium at $T\sim 2T_c$, especially in view of the small differences
observed.  

\subsection{Sixth and higher order}
\begin{figure}
\begin{center}
\includegraphics[scale=0.6]{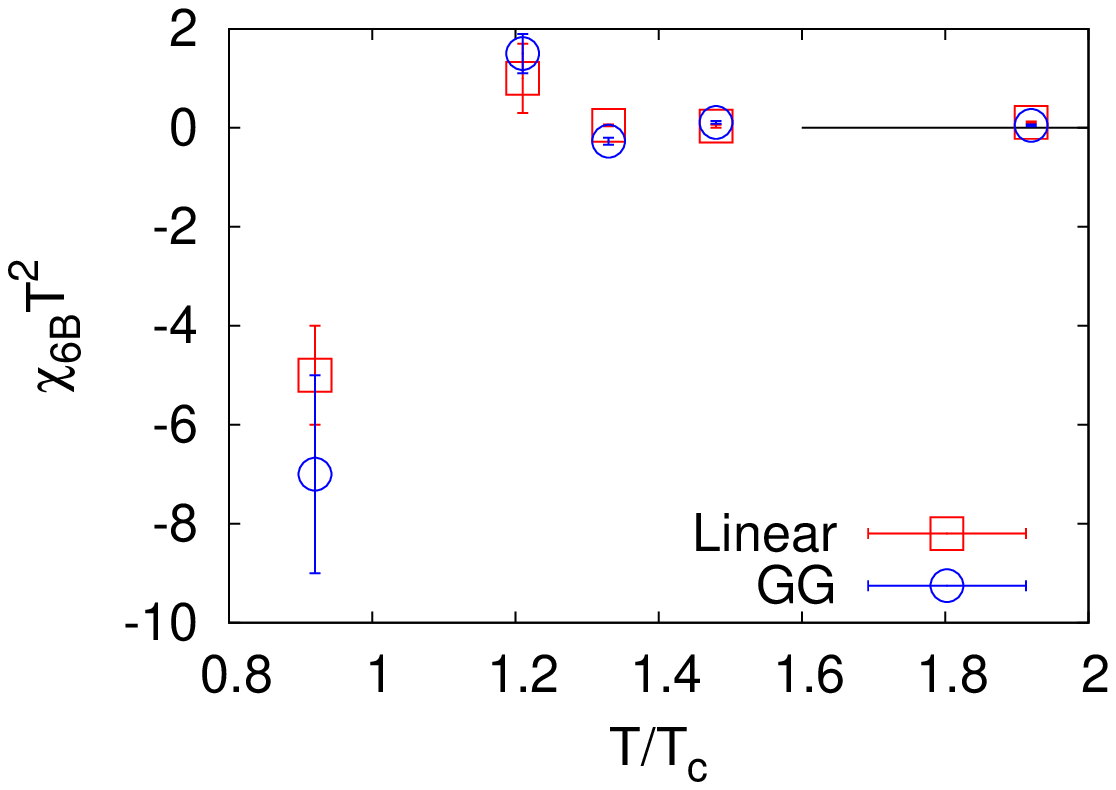}
\includegraphics[scale=0.6]{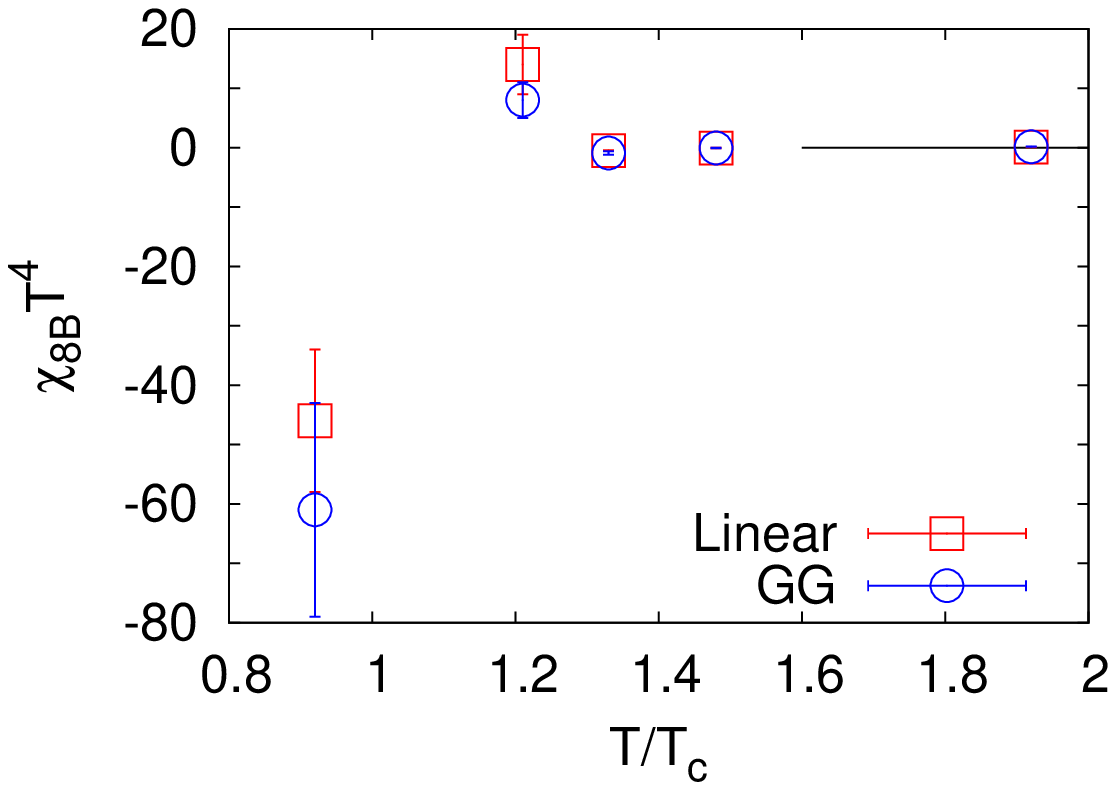}
\caption{The sixth(left panel) and eighth order(right panel) baryon number 
susceptibility as a function of $T/T_c$.}
\label{chi68B}
\end{center}
\end{figure}

For sixth order and above, we follow the lesson learnt from the free theory
and do not adapt any subtraction.   The sixth and the eighth order 
baryon number susceptibilities are displayed in Fig. (\ref{chi68B}).
We find again a good agreement with the GG results. At temperatures below
$T_c$ the larger error bars in the GG results compared to our results arise 
due to the presence of terms with varying sign in the former.  One has larger 
number of matrix inversions for computing such terms and one employs noisy 
estimators to determine them.  From the plots it is also evident that the
values of the higher order susceptibilities fall to zero rapidly for
$T>1.5~T_c$. The regime between $1.2-1.4~T_c$ may be still be sensitive to the
critical fluctuations as seen from the large deviations of the sixth and eighth
order fluctuations from the Stefan-Boltzmann values.  In the continuum, the
values of sixth and the higher order susceptibilities are zero for free
fermions and are finite in QCD only due to the interactions. To understand how
dominant are the interaction effects it is important to reduce lattice
artifacts and perform a continuum extrapolation of such quantities. Using the 
Dirac operator in Eq. (\ref{eqn:diracop}) these artifact effects are reduced as 
compared to the standard operator and also performing continuum extrapolation 
will be easier. 

\section{Ratios of susceptibilities and Radius of convergence}
The ratios of different QNS are sensitive indicators of the location of the
critical point and the extent of the critical region. For massless QCD, the
sixth and higher order susceptibilities should peak at $T_c$ with O(4) critical
exponents, implying indirectly the existence of the critical point.  Since our
input quark mass is quite large we should expect a crossover at $T_c$.  As
mentioned in Section \ref{basic formalism}, the ratios of susceptibilities are used to define
the radius of convergence and hence are important for determining the location
of the critical point.
 
Since the ratios are independent of the volume of the system, it was
suggested to use them in comparisons with data  a heavy ion collision
experiment where it is difficult to estimate the volume of the fireball formed.
Also such observables can be used for comparing lattice results at finite
volume with the experiments~\cite{sgupta}. The ratio of the fourth to the
second order baryon number susceptibility is related to the product of
kurtosis($\kappa$) and square of the variance($\sigma^2$) of the baryon number.
In the heavy ion collision experiments, the $\kappa\sigma^2$ of the net-proton
number is measured as a function of the center of mass energy of the colliding
heavy ion beams. A non-monotonic behavior of $\kappa\sigma^2$ as a function of
beam energy is a possible signature of the existence of the critical
point~\cite{gg3} and is currently being probed at the RHIC~\cite{mohanty}.  

\begin{figure}
\begin{center}
\includegraphics[scale=0.6]{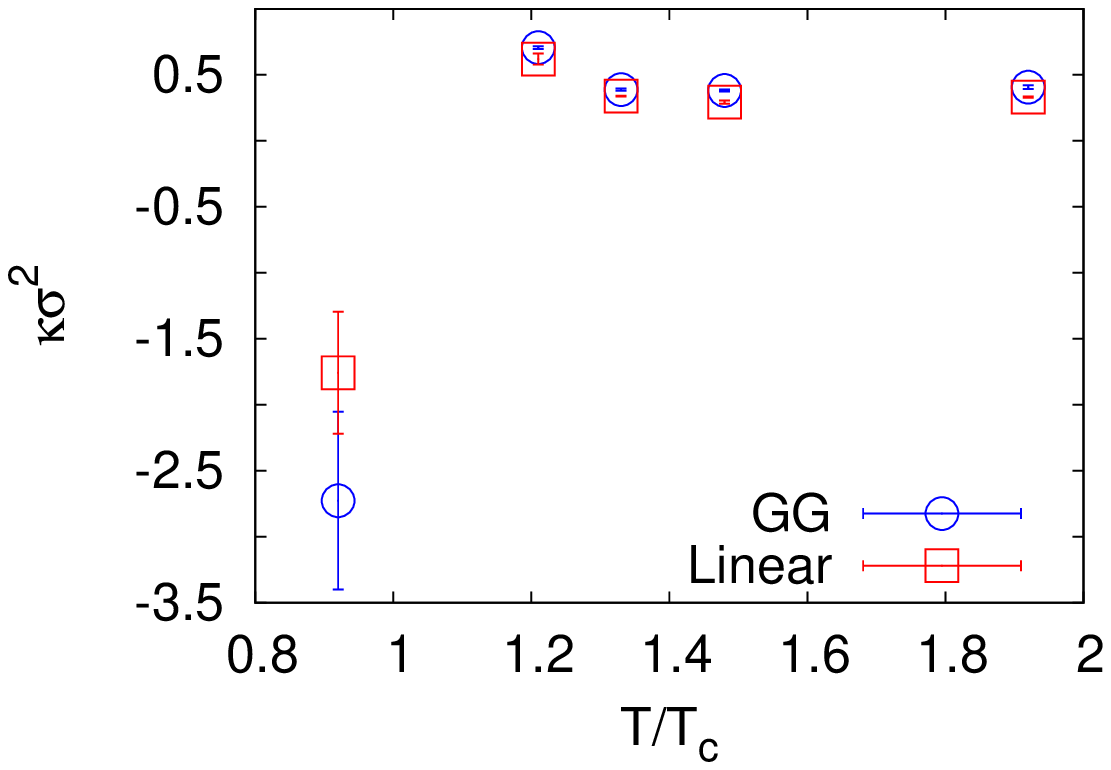}
\includegraphics[scale=0.6]{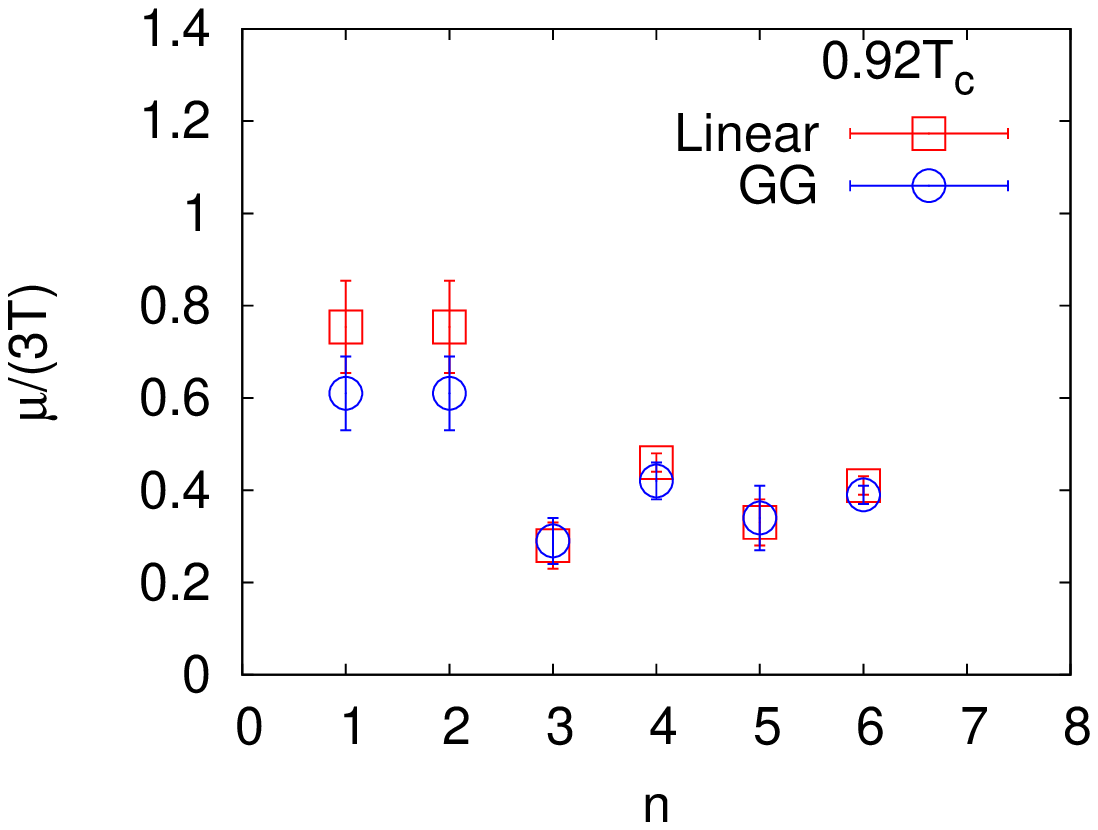}
\caption{The kurtosis(left panel) and different radii of convergence(right panel) as a function 
of $T/T_c$.}
\label{chiratio}
\end{center}
\end{figure}

We compute the quantity $K=\kappa\sigma^2$ and compare with the
value of the same computed from the susceptibility values in the
Ref.~\cite{gg2}. Note that these are still at zero chemical potential, and
are meant more for comparison of the two methods.
The results are displayed in the left panel of Fig. 
(\ref{chiratio}). A very good agreement is observed between the results in
these two different methods. This implies that the ratios of susceptibilities
are not very sensitive to the subtraction scheme used. We expect that the
radius of convergence estimates too would not be very sensitive to the
different subtraction schemes used. In order to verify this, 
different radii of convergence defined in Eq. (\ref{eqn:radc}) were computed at
$0.92 T_c$ using our results and compared with the corresponding GG
estimates.  The result of the comparison is shown in the right panel of Fig. 
(\ref{chiratio}). As expected the  radii of convergence computed using two
different subtraction schemes are roughly in agreement with each other. This is
promising as we can hope to now extend our analysis in the critical region with
susceptibilities at tenth order and beyond this way. We hope to be able to
check whether the current results on the location of the critical point are
changed in a a significant way with such estimates stemming from 
QNS beyond the eighth order.

\section{Conclusions}

Ever since it was recognized that the lattice free theory leads to $a^{-2}$
divergences in the continuum limit, mostly the exponential form for the
chemical potential term was used, as it eliminates the divergence for the
free theory.  Such an exponential form, however, seems not possible 
for the overlap quarks which have the same chiral properties as
the continuum.  Demanding chiral invariance for them even at finite density,
one obtains an overlap operator of the form linear in $\mu$.   Moreover, the 
linear form for any fermions, overlap or staggered, leads to simpler non-linear 
susceptibility expressions.  Absence of canceling terms, and requirement of fewer Dirac
matrix inversions make the linear form more suited for extension to higher
order non-linear susceptibilities needed to estimate the radius of 
convergence, or the location of the QCD critical point.

In this work we proposed a numerical scheme to remove the free theory 
artifacts inherent in the linear form.  In particular, subtractions were
introduced for the second and fourth order susceptibility which we
suggest should be evaluated for the free theory at zero temperature but the
same spatial volume.   As one sees from our Figs. ~(\ref{chi20},\ref{chi4B}), 
it works well.  
Indeed, the main difference from the popular exponential form stem from
the corresponding free theory due to the coarse finite spacing used at 
temperatures $T>1.5T_c$.
For the practically more relevant ratios, such as those in the left 
panel of Fig. (\ref{chiratio}) used as signature for critical point 
or in the right panel of Fig. (\ref{chiratio}), used for estimating the
radius of convergence, our method works as well as the exponential form
within errors.   It would be interesting to extend this work to higher
orders at the critical point, estimates of which are currently obtained 
with susceptibilities up to the eighth order.

\section{Acknowledgements}
The computations were performed on the Cray X1 of the Indian Lattice 
Gauge Theory Initiative(ILGTI) at TIFR, Mumbai. We thank Sourendu Gupta 
for many helpful discussions. 
We would like to thank ILGTI for providing the configurations used in 
this work. It is a pleasure to thank Kapil Ghadiali and Ajay Salve 
for technical support.
We gratefully acknowledge the financial support by the Alexander
von  Humboldt foundation and the kind hospitality of the theoretical physics
group of Bielefeld University, especially that of Frithjof Karsch and Helmut Satz. 
The research of RVG is partially supported by a J. C. Bose Fellowship from DST.
S.S. acknowledges Council for Scientific and Industrial Research for financial 
support during this work and thanks Saumen Datta for discussions.

\end{document}